\documentclass[acmsmall,nonacm]{acmart}

\usepackage{hyphenat}
\usepackage{xspace}

\usepackage{amsthm}
\theoremstyle{plain}

\newcommand{\abbrevStyle}[1]{#1}

\newcommand{\ie}{\abbrevStyle{i.e.}\xspace}
\newcommand{\eg}{\abbrevStyle{e.g.}\xspace}

\newcommand{\vs}{\abbrevStyle{vs.}\xspace}

\newcommand{\Secref}[1]{Sec.~\ref{#1}}

\newcommand{\Tabref}[1]{Table~\ref{#1}}
\newcommand{\Figref}[1]{Fig.~\ref{#1}}

\newcommand{\xhdr}[1]{\vspace{0.7mm}\noindent{{\bf #1.}}}

\newcommand{\textcite}[1]{\citeauthor{#1} \shortcite{#1}}

\newcommand{\hide}[1]{}

\makeatletter
\newcommand{\iffont}[2]{\ifthenelse{\equal{\f@family}{#1}}{#2}{}}
\makeatother

\iffont{ptm}{
  \usepackage{mathptmx}

  \DeclareSymbolFont{greek}{OML}{cmm}{m}{n}
  \DeclareMathSymbol{\alpha}{\mathalpha}{greek}{"0B}
  \DeclareMathSymbol{\beta}{\mathalpha}{greek}{"0C}
  \DeclareMathSymbol{\gamma}{\mathalpha}{greek}{"0D}
  \DeclareMathSymbol{\delta}{\mathalpha}{greek}{"0E}
  \DeclareMathSymbol{\epsilon}{\mathalpha}{greek}{"0F}
  \DeclareMathSymbol{\zeta}{\mathalpha}{greek}{"10}
  \DeclareMathSymbol{\eta}{\mathalpha}{greek}{"11}
  \DeclareMathSymbol{\theta}{\mathalpha}{greek}{"12}
  \DeclareMathSymbol{\iota}{\mathalpha}{greek}{"13}
  \DeclareMathSymbol{\kappa}{\mathalpha}{greek}{"14}
  \DeclareMathSymbol{\lambda}{\mathalpha}{greek}{"15}
  \DeclareMathSymbol{\mu}{\mathalpha}{greek}{"16}
  \DeclareMathSymbol{\nu}{\mathalpha}{greek}{"17}
  \DeclareMathSymbol{\xi}{\mathalpha}{greek}{"18}
  \DeclareMathSymbol{\pi}{\mathalpha}{greek}{"19}
  \DeclareMathSymbol{\rho}{\mathalpha}{greek}{"1A}
  \DeclareMathSymbol{\sigma}{\mathalpha}{greek}{"1B}
  \DeclareMathSymbol{\tau}{\mathalpha}{greek}{"1C}
  \DeclareMathSymbol{\upsilon}{\mathalpha}{greek}{"1D}
  \DeclareMathSymbol{\phi}{\mathalpha}{greek}{"1E}
  \DeclareMathSymbol{\chi}{\mathalpha}{greek}{"1F}
  \DeclareMathSymbol{\psi}{\mathalpha}{greek}{"20}
  \DeclareMathSymbol{\omega}{\mathalpha}{greek}{"21}
  \DeclareMathSymbol{\varepsilon}{\mathalpha}{greek}{"22}
  \DeclareMathSymbol{\vartheta}{\mathalpha}{greek}{"23}
  \DeclareMathSymbol{\varpi}{\mathalpha}{greek}{"24}
  \DeclareMathSymbol{\varrho}{\mathalpha}{greek}{"25}
  \DeclareMathSymbol{\varsigma}{\mathalpha}{greek}{"26}
  \DeclareMathSymbol{\varphi}{\mathalpha}{greek}{"27}
  \DeclareSymbolFont{otone}{OT1}{cmr}{m}{n}
  \DeclareMathSymbol{\Gamma}{\mathalpha}{otone}{0}
  \DeclareMathSymbol{\Delta}{\mathalpha}{otone}{1}
  \DeclareMathSymbol{\Theta}{\mathalpha}{otone}{2}
  \DeclareMathSymbol{\Lambda}{\mathalpha}{otone}{3}
  \DeclareMathSymbol{\Xi}{\mathalpha}{otone}{4}
  \DeclareMathSymbol{\Pi}{\mathalpha}{otone}{5}
  \DeclareMathSymbol{\Sigma}{\mathalpha}{otone}{6}
  \DeclareMathSymbol{\Upsilon}{\mathalpha}{otone}{7}
  \DeclareMathSymbol{\Phi}{\mathalpha}{otone}{8}
  \DeclareMathSymbol{\Psi}{\mathalpha}{otone}{9}
  \DeclareMathSymbol{\Omega}{\mathalpha}{otone}{10}
  \DeclareSymbolFont{syms}{OML}{cmm}{m}{it}
  \DeclareMathSymbol{\partial}{\mathord}{syms}{"40}
  \DeclareMathAlphabet{\mathbold}{OML}{cmm}{b}{it}
  \DeclareSymbolFont{largesymbols}{OMX}{cmex}{m}{n}

}
\usepackage{mathrsfs}
\usepackage{marginnote}
\usepackage{fancyvrb} 

\AtBeginDocument{%
  \providecommand\BibTeX{{
    \normalfont B\kern-0.5em{\scshape i\kern-0.25em b}\kern-0.8em\TeX}}}

\acmConference
\acmBooktitle



\newcommand{\NumSubreddits}{33\xspace}
\newcommand{\DurationExperiment}{63 days\xspace}

\newcommand{\NumUsers}{$97{,}616$\xspace}
\usepackage{subcaption}
\usepackage[breakable, theorems, skins]{tcolorbox}
\tcbset{enhanced}

\DeclareRobustCommand{\mybox}[2][gray!20]{%
\begin{tcolorbox}[   
        breakable,
        left=0pt,
        right=0pt,
        top=0pt,
        bottom=0pt,
        colback=#1,
        colframe=#1,
        width=\dimexpr\textwidth\relax, 
        enlarge left by=0mm,
        boxsep=1pt,
        arc=0pt,outer arc=0pt,
        ]
        #2
\end{tcolorbox}
}

\usepackage{tikz}
\usetikzlibrary{positioning, calc, shapes.geometric, shapes, shapes.multipart, arrows.meta, arrows, decorations.markings, external, trees}
\tikzset{
node distance=0.5cm, 
}
\tikzstyle{Arrow} = [
	thick, 
	decoration={
		markings,
		mark=at position 1 with {
			\arrow[thick]{latex}
			}
		}, 
	shorten >= 0.5pt, preaction = {decorate},
	]

\usepackage[utf8]{inputenc}
\usepackage{booktabs}
\usepackage{placeins}
\usepackage{graphicx}
\usepackage{caption}

\begin{document}

\title[Post Guidance for Online Communities]{Post Guidance for Online Communities}

\author{Manoel Horta Ribeiro}
\affiliation{%
  \institution{EPFL}
  \country{Lausanne, Switzerland}
}
\email{manoel.hortaribeiro@epfl.ch}

\author{Robert West}
\affiliation{%
  \institution{EPFL}
  \country{Lausanne, Switzerland}
}
\email{robert.west@epfl.ch}

\author{Ryan Lewis}
\affiliation{%
  \institution{Reddit}
  \country{San Francisco, US}
}
\email{ryan.lewis@reddit.com}

\author{Sanjay Kairam}
\authornote{Work completed while at Reddit. This author has since moved to OpenAI.\vspace{30mm}}
\affiliation{%
  \institution{Reddit}
  \country{San Francisco, US}
}
\email{sanjay.kairam@gmail.com}

\begin{abstract}
Effective content moderation in online communities is often a delicate balance between maintaining content quality and fostering user participation. In this paper, we introduce \textit{post guidance}, a novel approach to community moderation that proactively guides users' contributions using rules that trigger interventions as users draft a post to be submitted. For instance, rules can surface messages to users, prevent post submissions, or flag posted content for review. This uniquely \textit{community-specific}, \textit{proactive}, and \textit{user-centric} approach can increase adherence to rules without imposing additional burdens on moderators. We evaluate a version of Post Guidance implemented on Reddit, which enables the creation of rules based on both post content and account characteristics. Specifically, we conduct a large randomized experiment, capturing activity from \NumUsers posters in \NumSubreddits subreddits over \DurationExperiment.
We find that Post Guidance (1) increased the number of ``successful posts'' (posts not removed after 72 hours), (2) decreased moderators' workload in terms of manually-reviewed reports, (3) increased contribution quality, as measured by community engagement, and (4) had no impact on posters' own subsequent activity, within communities adopting the feature. Post Guidance on Reddit was similarly effective for community veterans and newcomers, with greater benefits in communities that used the feature more extensively. Our findings indicate that \textit{post guidance} represents a transformative approach to content moderation, embodying a paradigm that can be easily adapted to other platforms to improve online communities across the Web.
\end{abstract}

\setcopyright{acmlicensed}
\acmJournal{PACMHCI}
\acmYear{2025} \acmVolume{9} 
\acmNumber{2} 
\acmArticle{CSCW148} 
\acmMonth{4}\acmDOI{10.1145/3711046}

\ccsdesc[500]{Human-centered computing~Social networking sites}
\ccsdesc[500]{Human-centered computing~Empirical studies in collaborative and social computing}

\ccsdesc[500]{Human-centered computing~Empirical studies in collaborative and social computing}

\keywords{content moderation, online experiments, social media}

\maketitle

\begin{figure}
    \centering
    \includegraphics[width=.8\linewidth]{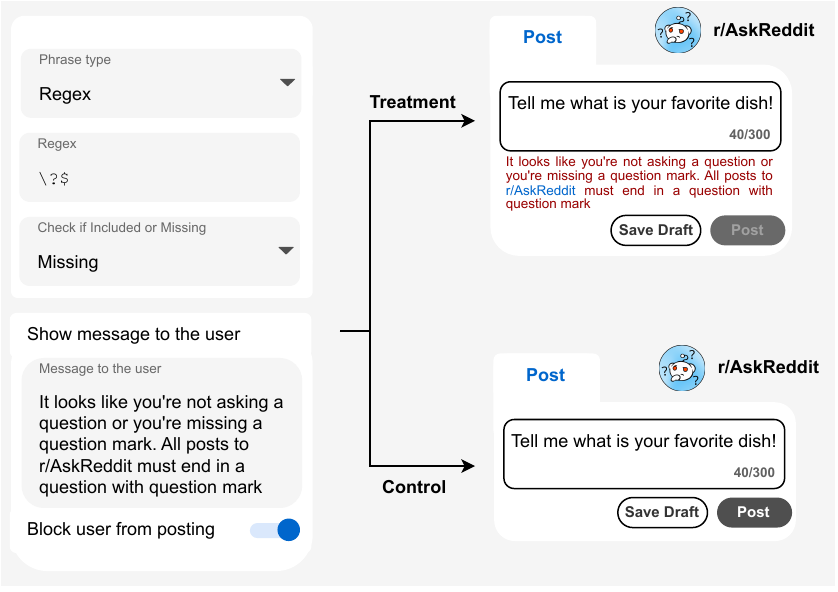}

    \caption{\textbf{How Post Guidance works.} When users try to write posts in a community, their contribution attempts are matched against a set of rules configured by the moderators of that community (left).
Posts can only be submitted if they fulfill these rules, \eg, the post in the image cannot be submitted as it does not end with a question mark.
Here, we present a user-level experiment ($n_{\text{users}}$=\NumUsers) where users were either exposed to this feature (`treatment') or not (`control') across \NumSubreddits communities on Reddit.}
    \label{fig:teaser}
\vspace{-3mm}
\end{figure}

\vspace{5mm}
\mybox{
\begin{center}
    {\textsc{This paper has been accepted at CSCW 2025. Please cite accordingly.}}
\end{center}}
\vspace{5mm}

\section{Introduction}

Rules and norms are crucial for online communities to thrive~\cite{kiesler2012regulating, fiesler2018reddit, chandrasekharan2018internet} and achieve their varied goals~\cite{weld2022makes, seering2023moderates, saha2020understanding, matthews2014goals}.
In \textit{r/relationshipadvice}, for instance, a community on Reddit where ``users can request others' opinions on a specific situation between two people,'' all posts must include the age and gender of the two referenced people using a stylized format, (\eg, `\textit{I [\textit{24M}] have an issue with my aunt [\textit{45F}]}').
Posts breaching this guideline will likely be removed either manually after review by the moderators of \textit{r/relationshipadvice} or automatically, through programmable moderation tooling, such as Reddit's `AutoModerator.' 

Moderators in communities such as \textit{r/relationshipadvice}, which have many contributions, must often decide whether to rely more heavily on automated tooling, which can scalably capture submissions that might deviate from the goals of the community, or to invest more time in manual moderation, making precise judgments about which content should or should not be accepted~\cite{seering2019moderator}. For instance, moderators might configure automated tools to remove all posts with keywords that correlate with some rule-breaking behavior; however, they then risk removing posts unfairly, leading disgruntled users to leave the community~\cite{jhaver2019human}. In contrast, moderators may consider each contribution manually and become overwhelmed by the large and constant task of evaluating content~\cite{schopke2022volunteer, matias2016civic}. Thus, there is a need for solutions that combine the scalability of automation with the nuanced understanding that humans can bring to evaluate content. 

A third promising direction focuses on shifting the knowledge and work associated with evaluating content from moderators to contributors, helping them better understand what content will or will not be accepted within the community. 
Prior research, for example, has explored simple `nudge'-like approaches, such as making rules salient~\cite{matias2016civic} or asking users to reconsider posts that may breach community guidelines~\cite{katsaros_reconsidering_2022}. These approaches are \textit{proactive} and \textit{user-centric}, guiding \textit{users} in how they should contribute to the community \textit{before} they break any rules, but also universal, in that this guidance does not adapt to the community or specific content. Other research has hypothesized that personalized guidance following moderation decisions may help encourage future contributions~\cite{jhaver2019does,jhaver2023bystanders}. These approaches are \textit{community-specific}, offering users the opportunity to improve their skills at evaluating what content will be accepted within a given community, but reactive and still requiring time-consuming manual intervention from moderators.

\subsection{Post Guidance} 
In this paper, we introduce \textit{post guidance}, a novel approach to moderating online communities that allows moderators to create rules that trigger interventions as users draft content in an online community. \textit{Post guidance} is a general paradigm for moderating user-generated content, which could be implemented in a variety of platforms hosting user-generated content, that match the following structure:
\begin{itemize}
    \item \textit{Users} draft and edit contributions within online platforms (\eg, posts, videos, or images). We refer to these contributions in the making as ``post drafts;''
    \item \textit{Platforms} implement ``interventions'' to prevent rule-breaking behavior. These can be customizable, \eg, sending custom messages. 
    \item \textit{Moderators} create ``conditions'' that, upon changes to the ``post draft'' (or attempts to submit it), may trigger ``interventions.''
\end{itemize}
Broadly, a \textit{post guidance} ``rule'' is a triplet $\langle$Intervention, Condition, Trigger$\rangle$. 
The intervention specifies \textit{what} the rule does.
The condition specifies \textit{when} the intervention is applied.
The trigger specifies \textit{where} the intervention is applied.

\xhdr{Implementing Post Guidance}
{In this paper, we evaluate the effectiveness of the \textit{post guidance} approach through a study of Reddit's recently-implemented Post Guidance feature~\cite{reddit2024postguidance}, conducted during an early experimental evaluation with a limited number of subreddits.}

{\Figref{fig:teaser} illustrates the {specific} implementation of Post Guidance on Reddit evaluated in this paper {for a typical rule.}
In \textit{r/AskReddit}, a prominent Q\&A community, all post titles must end with a question mark. Post Guidance can enforce this with a rule requiring the regular expression "\texttt{\textbackslash?\$}" to match every post title. 
Here, the intervention is to show a message and prevent the users from posting; 
the condition is that the post does not end with a question mark;
and the (implicitly defined) trigger determines that the rule is applied every time a user edits a post title.}

The specific implementation of Post Guidance considered only allows particular interventions, conditions, and triggers:
\begin{itemize}
    \item \textit{Interventions:} Rules may prevent users from posting and/or send a message to users while they are drafting the post.
    \item \textit{Conditions:} Conditions are specified with a regex expression. Moderators may require that the expression matches the content or not. Alternatively, users can list keywords and require that they are either present or absent from the content.
    \item \textit{Triggers:} Rules trigger when a user makes any change to the post's body and/or the post's title. This can be configured, although not shown in \Figref{fig:teaser}.
\end{itemize}
Several expansions to the above implementation are possible, as we discuss further in Section~\ref{sec:con}. Rules could flag posts or ban users from communities; they could have as conditions the output of machine learning classifiers; and they could trigger only when users try to submit their posts (and not when they are still drafting it).

Despite its simplicity, Post Guidance differs substantially from other content moderation interventions extant in large social media platforms; it is simultaneously \textit{community-specific}, \textit{proactive}, and \textit{user-centric}. This approach may improve adherence to rules and encourage contributions by educating individual users through a mechanism that reduces the effort required from moderators.
We provide additional examples of rules created using Post Guidance in Appendix~\ref{post_guidance:rules_examples}, and note that rules proposed varied from very generic and potentially applicable to various communities (e.g., preventing users from posting when URLs are title) to very specific (e.g., warning users that tech support is prohibited when they use words related to it).

\subsection{Hypotheses} 

Given this implementation of Post Guidance, we ask: \textit{can it improve the governance of online communities?} We explore this overarching research question via the four hypotheses below.

In many Reddit communities, moderators remove most or all rule-breaking posts, either through automated filtering or manual removal. The contextually relevant guidance provided by Post Guidance may lead some users to adapt their contributions to match the rules of the community, such that a subset of initially rule-breaking posts might avoid removal; thus, we hypothesize that:
\vspace{-1.5mm}
\mybox{
\textit{Post Guidance increases the number of non-removed contributions to communities adopting it.} \hfill (\textbf{H1}) 
}
\vspace{-1.5mm}
\noindent
Post Guidance will also prevent the submission of some subset of rule-breaking posts that would otherwise have required manual evaluation by moderators (either directly in the feed or within the AutoModerator queue); thus, we hypothesize that:
\vspace{-1.5mm}
\mybox{
\textit{Post Guidance decreases the workload for moderators in communities adopting it.} \hfill (\textbf{H2}) 
}
\vspace{-.5mm}

The first two hypotheses capture the ``direct effects'' of Post Guidance, but the feature may also indirectly shape communities. First, by ensuring that more posts align with the norms and goals of the community, it is possible that Post Guidance could contribute to an overall increase in a community-specified notion of post ``quality'', as measured by the degree to which other members engage with posts through votes and comments; thus, we hypothesize that:
\vspace{-1.5mm}
\mybox{\textit{Post Guidance increases the quality of contributions to communities adopting it.} \hfill (\textbf{H3}) 
}
\vspace{-1.5mm}
\noindent
Finally, it is possible that by helping more users successfully post and receive positive feedback from others within their communities, Post Guidance might help improve users' (especially newcomers') experience in and connection to those communities, as reflected in their frequency of visits and participation in the community; thus, we hypothesize that:
\vspace{-1.5mm}
\mybox{
\textit{Post Guidance increases users' engagement (other than posting) to communities adopting it.} \hfill (\textbf{H4}) 
}
\vspace{-1.5mm}

\subsection{Experimental Setup} 
To understand how Post Guidance shapes online communities, we conducted a large-scale field experiment with \NumSubreddits subreddits. Subreddits that opted to participate were onboarded onto the feature to ensure familiarity with the tool and the creation of rule-based post constraints. Over \DurationExperiment, \NumUsers users who started drafting at least one post in any of these subreddits were assigned randomly into treatment and control groups, enabling us to make within-subreddit comparisons. 
Using behavioral logs of the 28 days after enrollment, we calculated aggregated measures of (1) the users' posting activity, (2) moderation received on those posts, (3) community engagement with those posts, and (4) the users' overall engagement in the subreddit. This randomized setup allows us to analyze the causal effects of Post Guidance using simple regression analysis.

\subsection{Summary of Results} 

Our experiment offered strong evidence in favor of \textbf{H1}, \textbf{H2}, and \textbf{H3} and against \textbf{H4}. Post Guidance:

\begin{enumerate}
    \item[\textbf{H1}] Increased the number of ``successful'' contributions, which we operationalize as posts not removed 72 hours after being submitted ($+$5.8\% relative increase; $p$=0.03). Curiously, the feature reduced the number of posts started ($-$5.7\%; $p$<0.001) and submitted ($-$13\%; $p$<0.001), but posts created with Post Guidance were removed less often, which explains the increase.

    \item[\textbf{H2}] Decreased moderation workload. Users exposed to the feature had their posts reported less often ($-$9.4\%; $p$=0.001) and removed by the AutoModerator less often ($-$34.9\%; $p$<0.001). Note that moderators spend time reviewing both reported and automatically removed posts.
    Post Guidance did not significantly change the number of mod-removed posts ($+$2.7\%; $p$=0.236). 

    \item[\textbf{H3}] Increased the quality of contributions. As previously discussed, posts created with the feature were reported and removed less. 
    But they also received more comments ($+$28.6\%; $p$=0.004), screen views ($+$26.6\%; $p$=0.027), and more upvotes ($+$36.1\%; $p$=0.004), indicating that they abide by the subjective quality criteria in the community they were created.

    \item[\textbf{H4}] {Did not increase user participation. For users that engaged with the feature, we observed a small, not statistically significant \textit{decrease} in the number of days active  ($-$1.4\%; $p$=0.059), of votes   ($-$1.9\%; $p$=0.229) and of contributions ($-$2.0\%; $p$=0.223).}
\end{enumerate}

Subsequent secondary analyses revealed that the effect of Post Guidance was similar across Reddit veterans and newcomers. Additionally, subreddits that relied heavily on AutoModerator (\ie, automated, reactive content moderation) before the experiment and those that set up many Post Guidance rules saw the biggest increases in the number of `successful contributions.' This suggests that extensive use of the feature makes a difference and that going from reactive to proactive content moderation may yield considerable advantages to online communities.

\subsection{Implications and Future Directions} 
This study finds \textit{post guidance} to be effective as a scaleable and flexible content moderation paradigm, with the potential to improve user-generated content across the Web.
Though this approach is currently implemented for posts on Reddit, it would easily adapt to various contribution types and platforms, \eg, comments on social media, direct messages in instant messaging platforms, and wiki contributions in collaborative encyclopedias.
Moreover, the \textit{post guidance} paradigm can easily be extended using machine learning models that allow the feature to
(1)~handle a broader set of rules (e.g., ``Be kind'' and others that are hard to map to specific textual patterns);
(2)~handle multi-media content; and
(3)~provide even more personalized feedback. 
In that context, future work could extend this new paradigm ``breadth-wise,'' \ie, adapting it to other parts of the Web, or ``depth-wise,'' \ie, increasing the possibilities for rules and nudges and quality of the feedback.

\section{Background and Related Work}
\label{sec:rw}

We review previous work on content moderation on online platforms and provide background information about content moderation on Reddit.

\subsection{Content Moderation in Online Communities}
Online community platforms typically rely on multiple levels of content moderation to ensure not only that policy-violating content is removed, but also that communities publish contributions in line with their specific goals.

\xhdr{Platform-level and community-level moderation}
Commercial content moderation is often performed by workers who lack the cultural context associated with specific communities necessary to interpret content~\cite{roberts2016commercial}.
Recent work thus suggests that volunteer self-organized community moderation is a step into better-governed online spaces, as decisions would be more contextual and legitimate~\cite{seering2020reconsidering}.
From that viewpoint, scholars have argued that the extensive~\cite{li_all_2022} moderation work carried out by volunteers carries immense commercial~\cite{li_measuring_2022} and civic value~\cite{matias2016civic}, as it enables meaningful discourse to shop online.

In systems such as Reddit, Facebook Groups, or Discord, where users create their own online communities, each community often represents its own microcosm with different goals and idiosyncrasies~\cite{kairam2024founder}. On Reddit, for example, while some norms and expectations are shared across communities, previous research has found a long tail of explicit and implicit norms peculiar to some corners of the website~\cite{fiesler2018reddit,chandrasekharan2018internet}.
Given the highly contextual nature of the rules, a one-size-fits-all moderation approach is unfeasible, and, therefore, the work of enforcing community-specific rules is performed by volunteer moderators whose powers concern specific communities~\cite{edelson2023content,caplan2018content}. This represents a sharp contrast to networked platforms like Instagram or Twitter/X, where moderation follows centralized rules enforced by \textit{commercial} moderators~\cite{gillespie2018custodians}.

\xhdr{Distributed content moderation}
Many services enable online communities to control the visibility of content in a distributed fashion by allowing users to uprank and downrank content. Platforms like Facebook and Instagram use sophisticated ranking algorithms to generate feeds, often using implicit engagement metrics, e.g., predictions about how much time you will spend interacting with the post~\cite{meta_our_2023}. In contrast, community-driven platforms such as Reddit, Stack Overflow, and Slashdot often rely on explicit user feedback (\eg, upvotes) and the recency of posts~\cite{massanari2017gamergate} to determine what content is shown most saliently to users (this is often done with simple, deterministic algorithms, like Reddit's ``Hot'' algorithm).
Distributed content moderation is effective in filtering content that is accepted and appreciated by a community~\cite{lampe2004slash}; but previous research has found that it may propagate misinformation~\cite{gilbert2020run} and (further) marginalize minorities~\cite{das2021jol, massanari2017gamergate}.

\xhdr{Automated content moderation} 
In centralized online platforms, automated content moderation is concentrated in the hands of platforms, e.g., Horta Ribeiro et al. (2023) ~\cite{horta_ribeiro_automated_2023}  describe how Facebook uses various classifiers to detect and intervene upon rule-breaking content. Automated moderation in community-centered platforms, typically enables community moderators to set rules about what content is flagged or removed. Reddit's `AutoModerator,' for example, empowers \textit{moderators} to automate content moderation as they see fit. Given a configuration, the AutoModerator enacts moderation decisions like removing or flagging content based on the title or content of the submission or attributes associated with its creator, e.g., its karma.
Previous research has found that the AutoModerator reduces the workload and the emotional labor required of content moderation, but it is often too brittle and fails to capture nuances~\cite {jhaver2019human}.

\subsection{Understanding and Improving Content Moderation}
A growing body of work has measured the causal effect of existing content moderation practices through experimental and quasi-experimental research designs. At times, researchers have also proposed new content moderation practices to improve online platforms. This literature has informed the development of Post Guidance and its assessment.

\xhdr{Community or platform-level interventions}
Some work has focused on large-scale, platform-wide moderation interventions, e.g., banning popular influencers from Twitter or YouTube~\cite{jhaver_evaluating_2021,rauchfleisch_deplatforming_2021,klinenberg_does_2023}, limiting access or banning toxic online communities~\cite{chandrasekharan_quarantined_2022,trujillo_make_2022, trujillo_one_2023,chandrasekharan_you_2017, russo_spillover_2023, russo_understanding_2023, russo_stranger_2023, horta_ribeiro_platform_2021}, and even banning entire fringe social media platforms like Parler~\cite{horta_ribeiro_deplatforming_2023,kumarswamy_impact_2023}.
Overall, this work suggests platform-level interventions reduce activity and the capacity of impacted communities or interest groups to attract new members~\cite{jhaver_evaluating_2021,rauchfleisch_deplatforming_2021,klinenberg_does_2023,horta_ribeiro_platform_2021}. 
But this may come at the cost of communities creating smaller, more radical, and polarized communities~\cite{trujillo_make_2022,trujillo_one_2023,horta_ribeiro_platform_2021, ali_understanding_2021} that are oftentimes less public-facing~\cite{urman_what_2022}., e.g., Urman and Katz (2022) found that far-right Telegram groups experienced explosive growth coinciding with the banning of far-right actors on mainstream social media platforms~\cite{urman_what_2022}.

\xhdr{Fine-grained content moderation interventions} Although platform-level moderation interventions can shape our information ecosystem, most content moderation efforts concern micro-level decisions about millions of posts, comments, images, and videos. 
In that direction, previous work has analyzed the effect of removing content on various platforms~\cite{srinivasan_content_2019,horta_ribeiro_automated_2023,jimenez-duran_economics_2023,mitts_removal_2022}, banning users~\cite{ali_understanding_2021,mitts2021banned}, and labeling content that might be misleading or polarizing~\cite{bhuiyan_nudgecred_2021,clayton_real_2020,gao_label_2018,ling_learn_2023,zannettou_i_2021,porter_political_2022,mena_cleaning_2020,ecker_explicit_2010}.
Altogether, this body of work suggests that the effectiveness of various moderation strategies in shaping subsequent user behavior is mediated by their target and how the intervention is carried out. For example, Jhaver et al. (2019)~\cite{jhaver2019did} suggest that users are more likely to post again after having their content removed if they perceive the decision as fair. 
Likewise, Martel and Rand (2023), in reviewing the warning labels literature, indicate that they are particularly beneficial for decreasing the belief and spread of politically agreeable content~\cite{martel_misinformation_2023}.

\xhdr{Proactive content moderation}
So far, the moderation interventions discussed are reactive: they are enforced \textit{after} users or communities have broken platform rules. Yet, a growing body of research has focused on preventing users from breaking the rules or moderating content before it is widely shared~\cite{habib_act_2019,ribeiro_post_2022, schluger_proactive_2022, zhang_conversations_2018}.
In that direction, we highlight previous work by Chang et al. (2022), who, in an experiment, found that users became more civil when aided by a tool that indicated when conversations might be going awry~\cite{chang_thread_2022}.
{Closely related to \textit{proactive moderation} are access or participation controls~\cite{kiesler2012regulating}. For instance, Seering, Kraut, and Dabbish (2017) have found that streamers use ``chat modes'' on Twitch (settings which alter how spectators can participate) to shape audience behavior, (e.g., preventing spam)~\cite{seering_shaping_2017}.}

\xhdr{User-centric content moderation} Moderation interventions may burden moderators (being ``moderator-centric'', \ie, moderators are the ones removing posts) or users in the community (``user-centric'').
In the existing literature, the most noteworthy user-centric interventions entail making norms salient.
Bringing attention to the rules and expectations in online spaces has consistently improved user behavior in online platforms---no matter whether done reactively or proactively~\cite{tyler_social_2021,matias2019preventing,katsaros_reconsidering_2022,jhaver2019did}.
{For example, Katsaros et al. (2022) show that, for users who post tweets containing offensive content, asking them to reconsider decreased future creation of offensive content~\cite{katsaros_reconsidering_2022}.
Other work shows that even when norms are made salient \textit{after} content is removed, e.g., when ``removal explanations'' are provided to users, there is an improvement in subsequent user behavior~\cite{jhaver2019does}.}
Last, recent work has shown that providing explanations about removals in public~\cite{jhaver2023bystanders} or highlighting may improve user behavior online by itself~ \cite{wang_highlighting_2022}, suggesting that bystanders learn community norms and etiquette by example. 

\subsection{Content Moderation on Reddit}

Reddit comprises over 100${,}$000 active communities (sometimes called \textit{subreddits})~\cite{reddit2023api}, where users can contribute submissions, comment on others' submissions, and upvote or downvote others' comments and submissions. These communities, independently created and managed by users called \textit{moderators}, cover a broad range of interests and topics; as Reddit puts it, ``whether you're into breaking news, sports, TV fan theories, or a never-ending stream of the internet's cutest animals, there's a community on Reddit for you''~\cite{reddit_what_2023}.

Activity within communities must abide by Reddit's platform-level Content Policy~\cite{reddit_content_2023} and Moderator Code of Conduct~\cite{reddit2023moderator}, which are enforced by administrators (or \textit{admins}), who are Reddit employees, as well as the ``Reddiquette'' values and practices that broadly apply to the platform~\cite{reddit_reddiquette_2023}. Communities themselves create their own community-specific rules and norms, which are enforced by the community moderators~\cite{chandrasekharan2018internet, fiesler2018reddit}. Community members can engage in distributed moderation by upvoting or downvoting content, controlling its visibility within the feed on a specific community. Reddit supports automated content moderation via its AutoModerator system, used extensively across the service, particularly in large subreddits~\cite{basu_reddits_2019}, alongside a number of other automated systems for safe-guarding communities and redditors~\cite{reddit2024keeping}. 
We stress that while `automated,' AutoModerator is not user-centric, as it does not shift the burden of moderation onto users: moderators often have to review posts deleted or flagged by the tool, or to deal with complaints associated with removals enacted by it~\cite{jhaver2019human}.
Reddit also has a reputation system called `karma,' an integer calculated using upvote data representing how much a user's contributions have been appreciated within the website~\cite{reddit_what_2023}. 

\subsection{{Research Gap}}

We conclude our related work section by clarifying the research gap addressed by \textit{post guidance}.
Self-organized community moderation can produce well-governed online spaces where decisions are more legitimate and better contextualized~\cite{seering2020reconsidering}.
This paradigm is bound by the (usually unpaid) labor of community moderators~\cite{li_measuring_2022}, and therefore, interventions and systems that reduce their workload can help improve our online information ecosystem. 
Yet, existing systems and interventions meant to facilitate moderators' work often discourage user participation~\cite{wright2022automated} or create additional tasks that also consume moderators' time~\cite{jhaver2019human}.
In that context, \textit{post guidance} emerges as a complementary intervention that may improve adherence to rules and encourage contributions by educating individual users through a mechanism that reduces the effort required from moderators.
{We evaluate this paradigm here through a specific implementation of Reddit's Post Guidance feature.}

\begin{figure}
    \centering
    \includegraphics[scale=0.5]{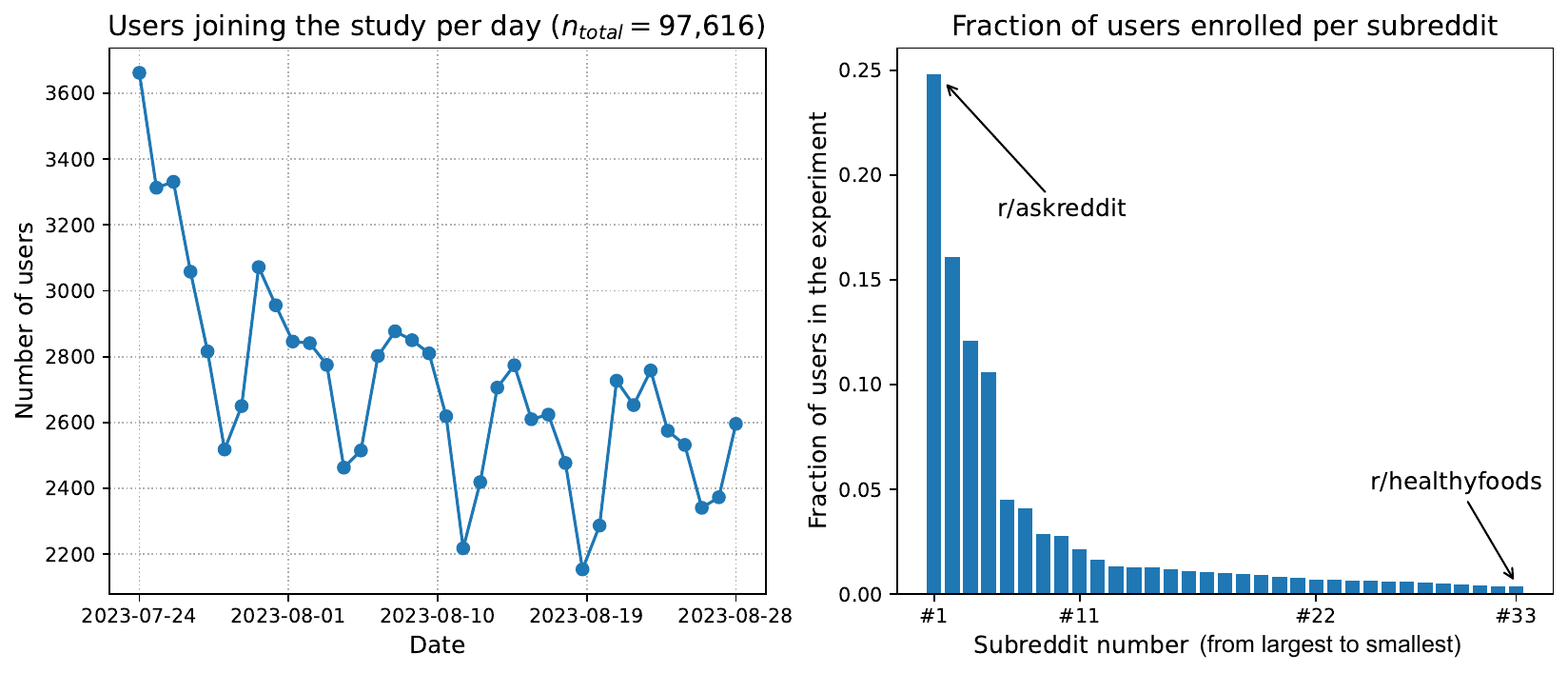}
    \caption{\textbf{Details about experiment enrollment.} On the left, we show the number of users enrolled in the experiment per day; there is seasonality (with lower enrollment during weekends) and a continuous drop in the number of enrolled users in the first few weeks (as users can only enroll once).
    On the right, we show the fraction of users in the experiment per subreddit considered; there is a long tail of subreddits with less than 1\% of users in the experiment. Note that subreddits are sorted by the number of distinct users that entered the posting interface during the study period (from largest to smallest).}
    \label{fig:enrollment}
    \vspace{-5mm}
\end{figure}

\section{Materials and Methods}

We describe a large ($n$=\NumUsers) randomized experiment to test the effectiveness of Post Guidance in improving community-level content moderation on Reddit. The experiment was conducted over a period of \DurationExperiment on \NumSubreddits subreddits. High-level descriptive statistics for these subreddits are provided in Appendix~\ref{appendix:communities-list}, and rule examples are shown in Appendix~\ref{post_guidance:rules_examples}.

\xhdr{Pre-experimental setup}
Since Post Guidance is a community-level moderation strategy, the critical pre-experimental setup step in the study was to onboard subreddits. 
Post Guidance only works if there are rules in place, and therefore, we needed to give moderators access to the tool and encourage them to adopt it as part of their day-to-day moderation operations. 
Subreddits either self-enrolled after an announcement in a private subreddit gathering moderators of large communities or were invited to participate by three Reddit employees, who helped moderators in these communities familiarize themselves with the tool and, in some cases, optimize their rules.
While enrollment to use the feature was continuous, we focused on the first \NumSubreddits subreddits that adopted the feature and created at least two post-guidance rules.

\begin{figure}[t]
\centering
\begin{minipage}[t]{\textwidth}
\begin{minipage}[t]{0.49\textwidth}
\subcaption{}
\label{fig:4a}
\includegraphics[scale=0.5]{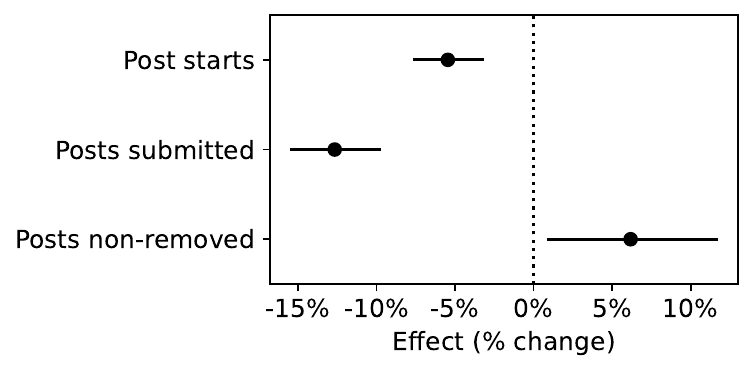}
\end{minipage}\hfill%
\begin{minipage}[t]{0.49\textwidth}
\footnotesize
\subcaption{}
\label{tab:tabxx1x}

\vspace{3.25mm}
\begin{tabular}{lllll}
\toprule
& \multicolumn{2}{l}{Control} & \multicolumn{2}{l}{Treatment} \\
 Posts (...) &       \# & \%$\Delta$ &         \# & \%$\Delta$ \\ \midrule
starts       &   85421 &        — &     80593 &        — \\
submitted   &   53500 &    37.4\% &     46522 &    42.3\% \\
non-removed &   23268 &    56.5\% &     24527 &    47.3\% \\
\midrule 
Total users & 48793 & & 48823 & \\
\bottomrule \\
\end{tabular}

\end{minipage}
\end{minipage}
\caption{
\textbf{Effect of Post Guidance on contribution success.} (a) Post guidance significantly reduces the number of non-removed posts, even though fewer posts are started and submitted; (b) We further detail the fraction of (potential) posts lost at each step of the creation pipeline, indicating the number ($\#$) and the percentage of post ``lost'' at each step (\%$\Delta$) for treatment and control groups.
}
\label{fig:step4}
\end{figure}

\xhdr{Assignment procedures} All individuals on desktop (excluding users of `old Reddit,' a pre-2018 version of the interface) who opened the interface to draft a post in one of the \NumSubreddits enrolled subreddits were enrolled in the experiment. They were randomly assigned to either treatment or control with equal probability (meaning we have roughly 49${,}$000 users in each condition; see \Figref{tab:tabxx1x} for the exact figures).%
\footnote{Note that this implies that larger subreddits (i.e., where more users open the `draft post' interface) have more users recruited to the experiment.}
Importantly, once a user is enrolled in a specific subreddit, we only consider the subsequent activity of that user in that subreddit, even though they will be sorted in the treatment or control group for all subreddits in the study; \eg, for a user who first opened the interface to draft a post on \textit{r/AskReddit} and was assigned to the treatment group, if they then went on to open the interface to draft a post on  \textit{r/healthyfoods}, they would also have Post Guidance enabled. 
However, we do not consider cross-community effects in the experiment at hand, as we find that the initial subreddits users visited when they were enrolled are responsible for most subsequent activity (\eg, 97\% of subsequent posts; 98\% of subsequent daily activity).

\begin{figure}
    \centering
    \includegraphics[scale=1.1]{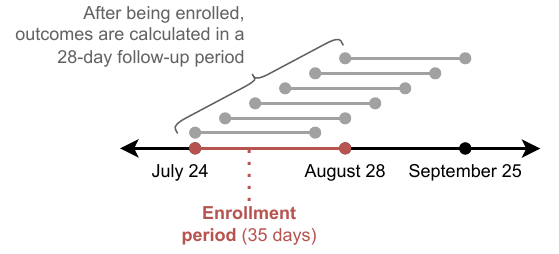}
    \caption{\textbf{Timeline of the experiment.} Users were enrolled in a 35-day period between 24 July 2023 and 28 August 2023. After enrolling, outcomes are calculated in a 28-day follow-up period. In the figure, we use gray lines to symbolize the tracking period of hypothetical users enrolling in different days.}
    \label{fig:timeline}
\end{figure}

\xhdr{Start date and follow-up} Enrollment started on 24 July 2023 and ended on 28 August 2023, and for each enrolled user, we considered a follow-up period of 28 days.
In other words, we track users for 28 days after they first attempt to create a post in one of the {\NumSubreddits} considered subreddits; e.g., if someone were enrolled on 28 August, we would track them until 25 September.
In total, \NumUsers users enrolled in the experiment. 
\Figref{fig:enrollment} depicts the number of enrolled users per day and the fraction of users enrolled per subreddit.
\Figref{fig:timeline} illustrates the experimental timeline; the enrollment period is shown with a red line along the main axis of the figure and grey lines above the main axis depict the tracking period of hypothetical users enrolling on different days.

\xhdr{Outcomes} We consider thirteen different outcomes, which we describe in detail in \Tabref{tab:outcomes}. 
In short, we consider variables related to the post creation flow (\eg, Post starts), content moderation (\eg AutoModerator removals), post engagement (\eg, Received comments), and user activity (\eg, Days active).
We tie each outcome to one of our research hypotheses, with the exception of `Number of reports,' which is associated with both \textbf{H2} and \textbf{H3}. We plot the distribution of the outcomes in \Figref{fig:dist} (at the end of the paper).
We note that when considering outcomes related to\textbf{ }H3, we jointly consider metrics associated with engagement (\eg, Rec. Comments) and the `desirability' of contributions (\eg, Num. Reports). This is in alignment with previous literature studying contribution/article quality in Wikipedia~\cite{halfaker2020ores}.

\begin{table}
\small
\caption{\textbf{Description of variables considered in the study.} We describe both main outcomes (\#1—\#13) and variables used to study the heterogeneity of the effect (A—D). For the former, we indicate the hypothesis they are tied to; for the latter, we indicate whether they were calculated at the user (User) or subreddit-level (SR).}
\label{tab:outcomes}
\begin{tabular}{lrp{7cm}r}
\toprule
\textbf{\#} & \textbf{Name} & \textbf{Description} & \textbf{Hyp./Kind}\\\midrule
1 & Post starts & Number of times the user has entered the post creation interface in their assigned community. & \textbf{H1} \\\midrule
2 & Post submitted & Number of times the user has submitted posts in their assigned community. & \textbf{H1}\\\midrule
3 & Post non-removed & Number of posts made by the user in their assigned community that were not removed in the 24 hours after they were submitted. & \textbf{H1} \\\midrule
4 & Automod removals & Number of times the AutoModerator has removed posts or comments from the user in their assigned community. & \textbf{H2}\\ \midrule
5 & Mod removals & Number of times a moderator has removed posts or comments from the user in their assigned community.  & \textbf{H2}\\\midrule
6 & Admin removals & Number of times an admin (a Reddit employee with Reddit-wide moderation capacities) has removed posts or comments from the user in their assigned community.& \textbf{H2}\\\midrule
7 & Num. reports & Number of times other users reported posts by the user in their assigned community.& \textbf{H2}, \textbf{H3}\\\midrule
8 & Rec. comments & Number of comments (from other users)  in posts made by the user in their assigned community. & \textbf{H3} \\\midrule
9 & Rec. screen views & Number of screen views (from other users) in posts made by the user in their assigned community. & \textbf{H3} \\\midrule
10 & Rec. upvotes &Number of upvotes (from other users) received in posts made by the user in their assigned community. &  \textbf{H3} \\ \midrule
11 & Days contributing & Numbers of days in the follow-up period (maximum 28) that the user has created a post or a comment in their assigned community. & \textbf{H4} \\\midrule
12 & Days voting & Numbers of days in the follow-up period (maximum 28) that the user has up or downvoted a post or comment in their assigned community. & \textbf{H4}  \\\midrule
13 & Days active & Numbers of days in the follow-up period (maximum 28) that the user visited their assigned community. &  \textbf{H4} \\\midrule
A & Newcomers & Whether the user visited the subreddit in the 90 days before enrolling in the experiment. & User \\\midrule
B & Low activity & Whether the user is in the bottom quartile of activity in the 90 days before enrolling in the experiment. & User \\\midrule
C & High rule count & Whether the subreddit created more than 7 rules in the 90 days before the start of the experiment. & SR  \\ \midrule
D & High automod-use & Whether the AutoModerator touched more than 8\% of posts in the 90 days before the start of the experiment. & SR
\\\bottomrule
\end{tabular}
\end{table}

\xhdr{Statistical analyses} 
Given that outcomes are heavy-tailed counts, we estimate the average treatment effect with a Poisson Regression model
\begin{equation}\label{eq:1}
    \log \mathrm{E}(Y|Z) = \alpha + \beta \cdot Z,
\end{equation}
where $Y$ is the outcome we care about, and $Z\in\{0,1\}$ is a binary indicator marking whether the user is in the treatment or the control group. Note that $\beta$ captures the log ratio between the treated and control groups, \ie, $\beta = \log \frac{\mathrm{E}(Y|Z=1)}{\mathrm{E}(Y|Z=0)}$.
Coefficients estimated with the Poisson Regression are \textit{consistent} and \textit{unbiased}, even if the dependent variable $Y$ is not Poisson-distributed~\cite{cameron2013regression}.
Yet, Poisson Regression assumes that $\mathrm{E}(Y) = \mathrm{Var}(Y)$, which creates problems in estimating standard errors.
We address this issue using a robust covariance matrix estimator (commonly known as \textit{HC0} or Huber estimator), see~\cite{palmer2007overdispersion}.

\xhdr{Heterogeneity of the effect} 
We study how the effect varies depending on the user's and the subreddit's characteristics.
To do so, we stratify the effect according to user\hyp{} and subreddit\hyp{}level variables.
For example, assuming a dummy variable $X \in \{0, 1\}$, we extend Eq.~\ref{eq:1} to
\begin{equation}\label{eq:2}
    \log \mathrm{E}(Y|Z,X) = \alpha +  \beta \cdot  Z + \eta \cdot X + \gamma \cdot Z \cdot X ,
\end{equation}
where $\gamma$, captures whether the effect is `stronger' when $X$=$1$; it captures the ratio between the relative effect in units where $X$=$1$ and units where $X$=$0$, \ie, $\gamma$=$\log \Huge[ \frac{\mathrm{E}(Y|Z=1,X=1)}{\mathrm{E}(Y|Z=0,X=1)} /
\frac{\mathrm{E}(Y|Z=1,X=0)}{\mathrm{E}(Y|Z=0,X=0)}\Huge]
$.
We consider stratifying the effect along four distinct covariates, depicted in \Tabref{tab:outcomes}, capturing whether the user was a newcomer (A) or highly active (B), and whether the subreddit implemented many Post Guidance rules (C) or used AutoModerator substantially before the start of the experiment (D).

\section{Results}

Our experiment provide strong evidence in favor of \textbf{H1}, \textbf{H2}, and \textbf{H3} and against \textbf{H4}.
We present our results in \Tabref{tab:tabres} and discuss them in further detail below.
{Note that Table 2 provides significance levels adjusted with a simple Bonferroni correction to help the reader understand the impact of multiple testing. For clarity's sake, we do not adjust the p-values for each result.}

\subsection{H1: Effect of Post Guidance on Contribution Success} 
Post Guidance has significantly increased the number of non-removed contributions ($+$5.8\% relative increase; $p$=0.03). This happens even though there was a significant \textit{decrease} in post starts ($-$5.7\%; $p$<0.001) and an even larger and significant decrease in the number of submitted posts ($-$13.0\%; $p$<0.001).
We further illustrate this finding in \Figref{fig:step4}, plotting the effect size; and showing the percentage of contributions ``lost'' in each step of the posting pipeline.
{In the control group, more users start ($85{,}421$) and submit ($53{,}500$) posts compared to the treatment group (start: $80{,}593$); submit: $46{,}522$). Yet, a larger fraction of submitted posts are removed from the platform in the control group ($56.5\%$; $23{,}268$ non-removed) compared to the treatment group ($47.3\%$; $24{,}527$ non-removed).}

\begin{table}[b]

\begin{minipage}[t]{\textwidth}
    \centering
\small
\caption{\textbf{Poisson regression results.} Standard errors were calculated using the Huber robust estimator. We report effect sizes and confidence intervals as relative changes, e.g., a $-$5.7\% effect means that treated units experienced a relative decrease of 5.7\% relative to control units. {We also report the significance levels adjusted with a simple Bonferroni Correction (\textbf{Adjusted} $\alpha$). For Hypothesis 1, there is only one meaningful statistical test (Posts non-removed).} Last, note that to obtain effects as relative change, we simply transform the estimated effects using the formula $(e^{\beta} - 1)$. Given that  $\beta = \log \frac{\mathrm{E}(Y|Z=1)}{\mathrm{E}(Y|Z=0)}$, we have $(e^{\beta} - 1)= \frac{\mathrm{E}(Y|Z=1) - \mathrm{E}(Y|Z=0)}{\mathrm{E}(Y|Z=0)}$.
}
\label{tab:tabres}
\begin{tabular}{lllllr}
\toprule

\textbf{\#} & \textbf{Outcome} &                           \textbf{Effect} & $p$-\textbf{value} & \textbf{Adjusted} $\alpha$ & \textbf{Hyp.}\\
\midrule
 1 & Post starts       &     -5.7\%; 95\% CI [-7.8\%, -3.5\%] &    <0.001 &   0.05 &   H1  \\ \midrule
 2 & Posts submitted   &  -13.0\%; 95\% CI [-15.8\%, -10.2\%] &    <0.001 &  0.05  & H1  \\\midrule
 3 & Posts non-removed &       5.8\%; 95\% CI [0.6\%, 11.2\%] &      0.03 &    0.05 &   H1  \\\midrule
 4 & Automod removals  &  -34.9\%; 95\% CI [-37.0\%, -32.8\%] &    <0.001 &    0.0125 &  H2  \\\midrule
 5 & Mod removals      &       2.7\%; 95\% CI [-1.7\%, 7.2\%] &     0.236 &    0.0125 &   H2  \\\midrule
 6 & Admin removals    &    -9.2\%; 95\% CI [-17.3\%, -0.4\%] &     0.042 &    0.0125 &   H2  \\\midrule
 7 & Num. reports      &    -9.4\%; 95\% CI [-14.4\%, -4.1\%] &     0.001 &   0.0125 & H2, H3  \\\midrule
 8 & Rec. comments     &      28.6\%; 95\% CI [8.2\%, 52.9\%] &     0.004 &     0.0125 &  H3  \\\midrule
 9 & Rec. screen views &      26.6\%; 95\% CI [2.8\%, 56.0\%] &     0.027 &     0.0125 &  H3  \\\midrule
10 & Rec. upvotes      &     36.1\%; 95\% CI [10.1\%, 68.1\%] &     0.004 &     0.0125 &  H3  \\\midrule
11 & Days contributing &      -2.0\%; 95\% CI [-5.2\%, 1.3\%] &     0.233 &     0.0166 &  H4  \\\midrule
12 & Days voting       &      -1.9\%; 95\% CI [-5.0\%, 1.2\%] &     0.229 &      0.0166 &  H4  \\\midrule
13 & Days active       &      -1.4\%; 95\% CI [-2.9\%, 0.1\%] &     0.059 &     0.0166 &   H4  \\
\bottomrule 
\end{tabular}
\end{minipage}

\end{table}

\subsection{H2: Effect of Post Guidance on Content Moderation} 
\label{ssec:h2}
Contributions created using Post Guidance are more likely to remain in the platform because they are less likely to be removed. Users in the treatment condition had their posts removed less often by the AutoModerator ($-$34.9\%; $p$<0.001) 
and by Reddit administrators ($-$9.2\%; $p$=0.03), besides being reported less often ($-$9.4\%; $p$=0.001). Reviewing reports and AutoModerator removals is a substantial content moderation task, which implies that  Post Guidance decreased the moderation workload. 
We did not find a significant decrease in removals by moderators.
We conjecture that Post Guidance may, at the same time, (1) decrease rule-breaking posts from being created --- \eg, users might see that what they want to Post is not allowed and give up posting altogether; and (2) allow users to skirt rules, creating rule-breaking posts that would otherwise be caught by the AutoModerator --- \eg, upon seeing that their post break the rules, users might slightly alter their contribution in a way that leads to avoiding the AutoModerator, since the rules for Post Guidance and the AutoModerator are often similar.
To explore this further, we re-ran the analysis, considering only users who submitted a post in the follow-up period, a scenario that isolates the second mechanism outlined above. 
We find that users in the treated group who submitted a post were significantly more likely to have their posts removed by moderators (17.7\%; $p$<0.001), indicating that the two mechanisms conjectured above are at play (given that the overall effect is null).

\subsection{H3: Effect of Post Guidance on contribution quality} 
\label{ssec:h3}
Posts from users in the treatment group received more comments (28.6\%, $p$=0.004), screen views (26.6\%, $p$=0.027), and upvotes (36.1\%, $p$=0.004), suggesting they are overall better contributions.
In addition, as we previously mentioned, they were reported less often  ($-9.4\%$; $p$=0.001), which can also act as a proxy for very low-quality posts.
Note that posts created with \vs\ without Post Guidance co-existed in the subreddits during the experiment; these posts ``compete'' for users' upvotes, screenviews, and comments. Assuming that Post Guidance increases the quality of posts, it could be that engagement levels do not significantly increase as much as we see here once the feature is rolled out for all posts within a subreddit, as twice as many posts would increase in quality and thus relative engagement across all posts might decrease.

\subsection{H4: Effect of Post Guidance on user engagement} 
\label{ssec:h4}
Last, we find small and statistically insignificant effects when considering outcomes related to user engagement in their assigned subreddit. This contradicts our hypothesis that Post Guidance would increase user involvement.
Similar to in \textbf{H2}, we conjecture that two simultaneous effects may be at play here. On the one hand, Post Guidance increases the number of ``successful'' contributions, which could increase subsequent engagement. However, at the same time, Post Guidance raises the bar for users to participate in the community. To explore this further, we re-ran the analysis, again considering only users who submitted a post in the follow-up period. We find that, indeed, when considering this population, users exposed to Post Guidance experienced significant increases in subsequent engagement (Days contributing: 29.5\%; Days voting: 25.9\%; Days active 14.0\%; $p$<0.001).

\begin{figure}
    \centering
    \includegraphics[scale=0.5]{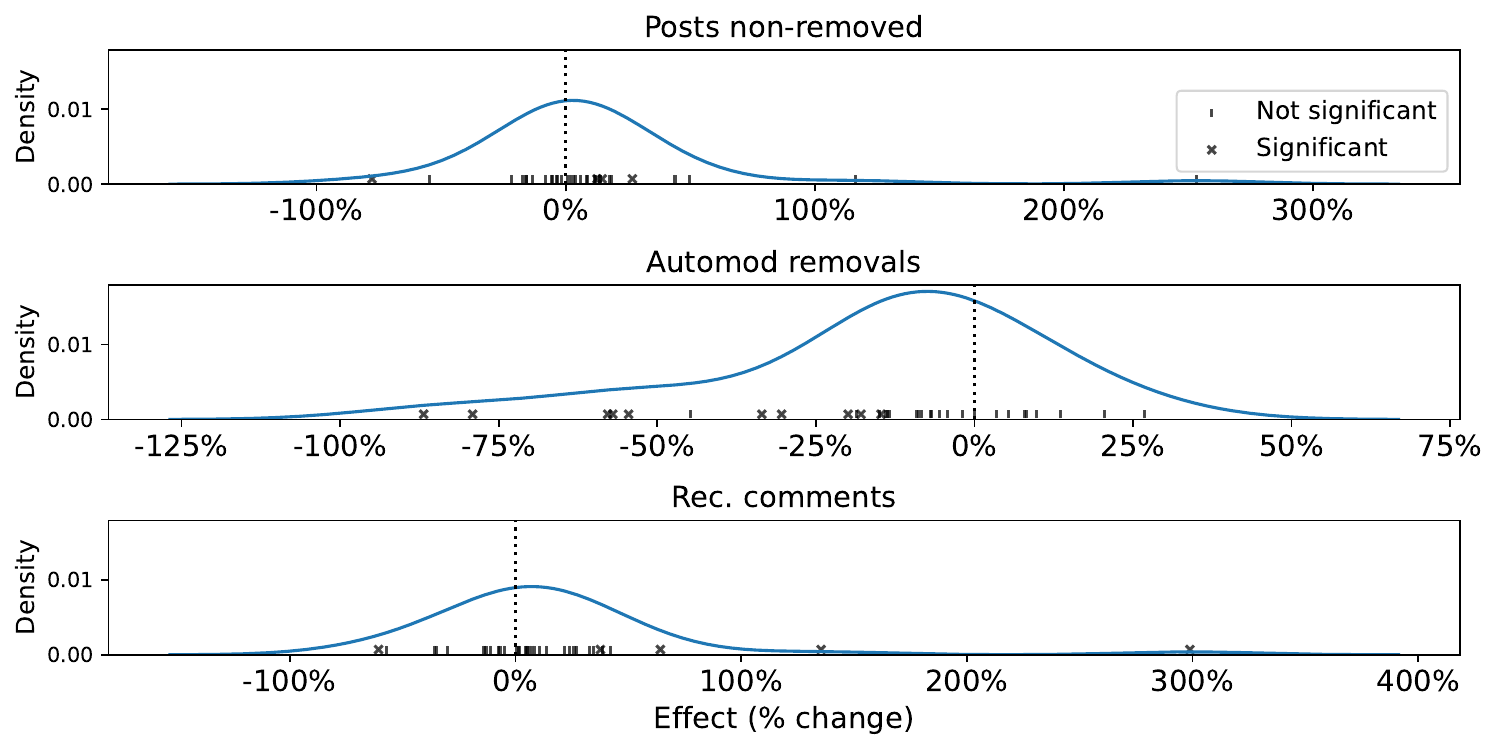}
    \caption{\textbf{Depiction of the heterogeneity of the effect.} For three of the outcomes considered (see \Tabref{tab:outcomes}), we show the distribution of subreddit-level effects with a kernel density estimate (KDE) plot. Close to the $x$-axis, we plot the actual effect sizes observed for each community using different markers for significant ($\times$) and not significant ($|$) subreddit-level effects. Note that effect sizes vary widely, going from negative to positive across all four outcomes (although, in many cases, the estimated effects are not statistically significant considering only one subreddit).
    To obtain the significance of the effects on a subreddit level, we repeatedly fit the Poisson Regression model depicted in Eq.~\ref{eq:1} to the data of each of the \NumSubreddits subreddits considered. The significance reported in \Figref{fig:het} is associated with the $p$-value of coefficients $\beta$ associated with each regression.
    }
    \label{fig:het}
\end{figure}

\begin{figure}[t]
    \centering
    \includegraphics[scale=0.5]{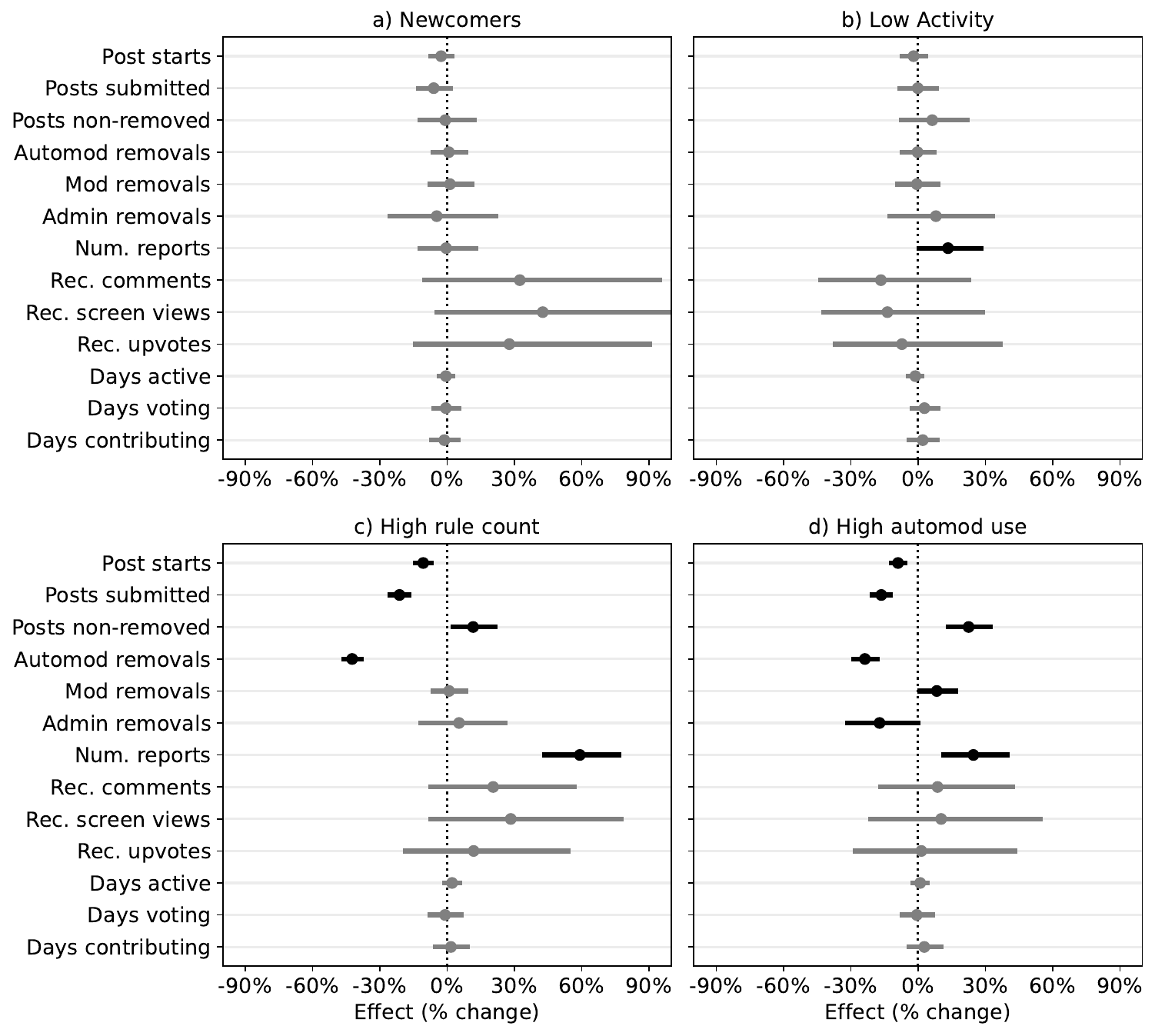}
    \caption{\textbf{Poisson regression results for effect heterogeneity.} Recall that estimates should be interpreted as measuring whether the effect is amplified or attenuated given specific conditions. For each binary variable $X$ (one per plot), we show the ratio between the effect between for users where this variable equals one and users where this variable equals zero. }
    \label{fig:het2}
\end{figure}

\subsection{Heterogeneity of the effect}
\label{ssec:ehet}
Not all communities are impacted by the Post Guidance in the same way. This is illustrated in \Figref{fig:het}, where we show that effect sizes vary widely across communities.
To further explore when Post Guidance is effective, we decompose the effect size into five components: one ``baseline'' effect and four ``interaction'' effects associated with subreddit and user characteristics (see \Tabref{tab:outcomes}); see Equation ~\ref{eq:2}. Recall that we can interpret the coefficients associated with the interaction ($\gamma$) as whether the effect is stronger or weaker when users or subreddits have specific characteristics. 

\xhdr{Newcomers}
We operationalize newcomers as users who, in the 90 days before enrolling in the experiment, did not visit their assigned community (54\% of users).
Surprisingly, we find that the effect for Post Guidance was not significantly different for newcomers when compared to other users for any of the 13 outcome variables considered (see \Figref{fig:het2}a). This indicates that people who are somewhat familiar with a community benefited equally from the feature compared to those who were not.

\xhdr{Low activity}
``Newcomer,'' as defined above, is a community-specific concept. A user might be a newcomer to r/AskReddit and, at the same time, be active in other Reddit communities.
We additionally consider users with `low activity' in the entirety of Reddit, operationalized as users with three or fewer votes on Reddit in the 90 days before enrolling in the experiment (50\% of users).
Results here are very similar to what we found for newcomers (see \Figref{fig:het2}b), with the sole exception of reports, which increased significantly more for these users relative to the others (relative increase of 12.5\%; $p$=0.04). 
We did not manage to hypothesize a credible reason for this observed effect (which could be a Type 1 error).

\xhdr{High rule count and AutoModerator use}
Subreddits were free to choose to which extent they adopted Post Guidance, and the number of rules in the \NumSubreddits considered subreddits ranged from 2 to 32.
{To contrast extensive and intensive Post Guidance use, we split our communities into two groups: those with a high rule count (more than seven rules; 46\% of users) and those with a low rule count (seven or fewer rules).
Further, we note that the functionalities of the AutoModerator, another automated content moderation tool at Reddit, overlap with Post Guidance.
For example, a community could either configure the AutoModerator to remove posts containing a specific keyword or configure Post Guidance to prevent posts from this keyword from being submitted.
Thus, we also split out communities into those with low AutoModerator use (less than 8\% of posts are touched by the AutoModerator; 52\% of the communities; 17 out of 33) and those with high AutoModerator use (more than 8\% of posts are touched the AutoModerator).

We find that the effect of Post Guidance differs in communities with high rule count and AutoModerator usage. 
Notably, we found a significant increase in the number of non-removed posts (11.0\% for high rule-count, $p$=0.03; 20.3\% for high AutoModerator removals, $p$=0.03);
a significant decrease in AutoModerator removals ($-$55.0\%, $p<$0.001; 27.0\%, $p<$0.001);
and a significant increase in the number of reported posts (46.5\%, $p<$0.001; 22.0\%, $p<$0.001).
Here, we attribute the significant increase in reported posts to the associated increase in non-removed posts. It may be that Post Guidance creates more `borderline' posts that are not considered rule-breaking by moderators, but that are perceived as so by users of the community.
Interestingly, we also find a significant increase in the number of posts removed by moderators in communities with high AutoModerator use (8.0\%, $p$=0.03). This may indicate that rules ported in this community from AutoModerator to Post Guidance allowed ``bad-faith'' rule-breaking users to try to skirt the rules (we discuss this further in \Secref{sec:con}).

\section{Discussion and Conclusion}
\label{sec:con}
Here, we propose Post Guidance, a novel approach to community moderation.
We show with a large-scale experiment that Post Guidance:
increased the number of non-removed contributions to communities adopting it \textbf{(H1)};
decreased moderator workload \textbf{(H2)}; and
increased the quality of contributions~\textbf{(H3)}. 
Interestingly, Post Guidance did not increase user engagement in the communities adopting it \textbf{(H4)}.
Our analyses also allow us to hypothesize how Post Guidance shapes online communities. 
For example, we find that it \textit{decreases} the number of posts submitted, but those that are submitted are less likely to be removed.

The effectiveness of Post Guidance was similar among more experienced users and newcomers (to Reddit and specific communities).
Yet, its effect varied across communities; those that saw the largest increases in the number of non-removed posts were the ones that
(1)~set up many Post Guidance rules; and
(2)~frequently used the AutoModerator feature before implementing Post Guidance.
This suggests a ``dose--response'' relationship between the changes in the community and the extent to which they use Post Guidance, as communities that set up more Post Guidance rules are bound to prevent more rule-breaking content, and as communities that heavily used the AutoModerator feature were likely to have rules that were easy to enforce with Post Guidance.
{Overall, we argue that Post Guidance can help improve the governance of online communities.
}

\xhdr{Design friction} 
While online platforms typically aim to simplify participation, Post Guidance is an example of `participation friction,' adding extra hurdles with the goal of ensuring that more submitted posts are successful. 
\Figref{fig:step4} supports this interpretation of Post Guidance as adding (positive) design friction, showing that fewer posts are submitted but that, in general, more posts survive in the community when the feature is enabled.
The friction added by Post Guidance may help explain our findings regarding \textbf{H4}. On the one hand, the feature increases subsequent user engagement for those users who are successful, similar to findings by Srinivasan (2023)~\cite{srinivasan2023paying}. On the other hand, the feature may `backfire' and discourage users whose posts do not follow the community's rules. Often, when a post or comment is removed \textit{after} submission, users are not even aware~\cite{jhaver2019does}; these users may continue to engage as if they had a successful post. By creating friction at the time of submission, Post Guidance may discourage this subsequent engagement. We find evidence supporting this interpretation in \Secref{ssec:h4}. 
While the overall result is null, if we limit our analysis to those who ended up submitting their posts, we find a strong positive effect for users exposed to the feature.

\xhdr{Good \vs bad faith rule breaking} 
Rule-breaking behavior can be broadly split between ``good-faith'' rule breaking, when users do not adhere to community norms because they do not know them,  and ``bad-faith'' rule breaking, when users are well aware of the rules but break them regardless. The iteration of Post Guidance studied here is mostly effective against good-faith rule-breaking, as it relies largely on regex patterns that can often be circumvented.
In contrast, this current iteration may facilitate bad-faith rule-breaking by helping users skirt the rules  (see \Secref{ssec:h2}). For example, a user may discover via Post Guidance that a specific word is forbidden in a community and then proceed to `fuzz' the word using punctuation to circumvent the rule. While the overall effect of Post Guidance, as studied here, is still positive, this balance can be further shifted in the future through the addition of machine-learning-powered evaluations of content, which go beyond keyword or regex matching and reduce the ability of bad-faith actors to circumvent community rules.

\xhdr{Proactive \vs reactive moderation}
The introduction of Post Guidance enables moderators to moderate content both proactively (Post Guidance) and reactively (using AutoModerator).
Interestingly, in \Secref{ssec:ehet}, we find that subreddits that relied heavily on automated \textit{reactive} moderation (high AutoModerator use) had significantly more non-removed posts and more manually removed posts than those that did not.
These differences could be explained by the overlapping functionalities between AutoModerator and Post Guidance: if subreddits already had meaningful AutoModerator rules (that triggered in many posts), it may have been particularly easy to `port' these meaningful rules to the new Post Guidance feature.
However, it may also be that these subreddits experienced more dramatic changes because they were the communities where a substantial chunk of automated \textit{reactive} moderation turned \textit{proactive}.
Last, it is important to stress that proactive and reactive moderation are complementary paradigms.
On the one hand, AutoModerator is better suited to implement regex rules to ``bad-faith'' rule-breaking (\eg, spam), as adversarial agents will likely exploit the Post Guidance interface to create rule-breaking posts. On the other hand, Post Guidance is better suited to educate users who want to conform to community guidelines.

\xhdr{User-centricity} 
An interesting way to examine content moderation features is to ask: who is burdened by it? 
In the case of a simple `delete button'  that reactively removes rule-breaking content, the answer is simple: it burdens moderators who will use it.
But even proactive content moderation features may burden content moderators; for instance, Horta Ribeiro et al.~\cite{ribeiro_post_2022} studied ``Post Approvals,'' a feature used in Facebook groups where every post has to be manually approved by moderators before landing in the groups' feed. Post Approvals effectively reduced low-quality posts but led moderators to create around-the-clock shifts to handle the demand of highly active communities~\cite{abidin2021subtle}. 
In contrast to these moderation practices that are ``moderator-centric,'' in the sense that they add work to moderators, Post Guidance is ``user-centric,'' \ie, it burdens users.
The work at hand, as well as previous work using nudges~\cite{matias2019preventing}, suggest that user-centric approaches are effective.

\xhdr{Contribution} This work meaningfully expands the growing literature on improving online spaces with fine-grained moderation interventions~\cite[\textit{inter alia}]{horta_ribeiro_automated_2023,jhaver2019did,bhuiyan_nudgecred_2021,srinivasan_content_2019} by 
1) proposing a new paradigm for user-centric, proactive, community-specific interventions; 
2) implementing said paradigm on Reddit; and 
3) comprehensively evaluating the impact of the implementation with a large randomized experiment in a major social media platform.
Our results suggest Post Guidance brings something unique to the table: it decreases rule-breaking behavior without creating new tasks for moderators (unlike interventions like Post Approvals~\cite{ribeiro_post_2022} or AutoMod~\cite{jhaver2019human}) and transcends `nudge'-like approaches~\cite{matias2016civic,bhuiyan_nudgecred_2021} by allowing moderators to tailor interventions to the needs of their communities.

\subsection{Limitations}
We conduct a large-scale randomized experiment "in the wild" on Reddit, meaning that our key findings (\ie, answers to \textbf{H1}--\textbf{H4}) have high internal and external validity. We note that the experiment was conducted only on the Web browser version of Reddit and not in the mobile app. Mobile users often differ in various ways from desktop users~\cite{antoun2015internet}, e.g.,  demographics, and generalizing the effects observed in mobile should be done with caution. This concern is partially addressed by our analyses of the heterogeneity of the effect, as we show that the effect is robust to user characteristics, \eg, the intervention works similarly for newcomers and veterans, as well as low and high-activity users.

Our secondary analyses have limitations worth discussing. First, in \Secref{ssec:h2} and \Secref{ssec:h4}, we conduct an analysis considering only users who submitted posts. This can yield biased results because there may be confounders that cause both users to submit posts and the other outcomes. Nonetheless, we argue that this analysis, albeit imperfect, provides us with further insight into the tradeoffs involved with Post Guidance, and we were not able to think of any particular confounder that would hinder the conclusions drawn in \Secref{ssec:h2} and \Secref{ssec:h4}. Second, when we conduct the analyses on the heterogeneity of the effect (\Secref{ssec:ehet}), we stress that we cannot attribute ``cause-and-effect'' interpretations to how the considered variables modify the effect. For example, there could be other features that cause both effect modification and lead subreddits to have a high rule count. Therefore, results on the heterogeneity of the effect should be interpreted as descriptive, \ie, how the effect differs for users with different characteristics.

\subsection{Implications and Future Work}
Post Guidance is a promising content moderation strategy that can improve online communities while reducing the moderation workload. More broadly, we believe that research evaluating the effect and nuances around content moderation (like this paper) can help improve the public debate around online platforms, making the available toolkit of interventions more transparent to stakeholders in academia, industry, and government.

\xhdr{Engaging users} The \textit{user-centric} aspect of \textit{post guidance} refers to the ways in which the overall approach requires users to actively engage with the rules of the community in order to contribute. While this study focused on the immediate effects, future work could explore how engaging users in this way could shape their relationships with their community. Prior work has shown how resolving ambiguity around specific applications of community rules can help moderators to iteratively build an understanding of a community's goals~\cite{cullen2022practicing}; shifting this effort towards users could confer similar benefits. Future work might also explore whether stronger engagement with the rules and goals of a community leads to the formation of stronger feelings of attachment to the community.

\xhdr{Adapting to Post Guidance} While we evaluate one specific implementation of Post Guidance in this work, the paradigm itself represents a tool that can flexibly be used by moderators alongside other tools. In our exploration of the heterogeneity of the effect, we find that Post Guidance was particularly efficient in increasing the number of non-removed posts in certain communities. Additionally, we find that the effectiveness of Post Guidance varied along with the use of existing moderation tools, such as Automoderator. Future work could explore how Post Guidance is, and could most effectively be, integrated by community moderators into a broader set of strategies and tools over time and adapted as the goals and needs of a community evolve, including shifting the roles and activities that moderators perform within their communities~\cite{seering2023moderates}.

\xhdr{Going beyond Reddit} 
Post Guidance could be used on other platforms hosting online communities, such as Discord and Facebook Groups. 
Adapting the \textit{post guidance} approach to the particularities of each platform could be an interesting venue for future work and would help improve online communities across the Web. 
We highlight that a proactive moderation feature like Post Guidance could be particularly interesting for Wikipedia, a large online community centered around building an encyclopedia. Wikipedia is one of the internet's greatest ``public goods,'' and struggles with retaining newcomers, especially since edits must follow a strict set of rules, and potential Wikipedians often give up after having their edits reverted~\cite{morgan2013tea,halfaker2011don}.
We conjecture that an intervention like Post Guidance could diminish these negative experiences and increase contribution to the world's largest encyclopedia.

\xhdr{Going beyond regex}
Post Guidance uses only simple rules programmable with regex. But with more complicated models, \eg, Large Language (multimodal) Models, one could imagine creating a more flexible version of Post Guidance. Some rules can be very easily enforced using regex (\eg, all posts must end with a question mark), whereas other can't (\eg, ``be kind,'' ``only post pictures of cats'').
More complex models could, therefore, increase the range of rules enforced by the community, allowing subreddits to deliver personalized prompts to a larger variety of scenarios. That being said, a big challenge in that direction is the loss of transparency --- although limited, a great virtue of the pattern-matching approach is that it gives community leaders very fine-grained control over automated moderation, which would perhaps be lost with more complex models.

\subsection{Ethical Considerations}
All communities involved in this experiment sought out and consented to participate. 
All data used in the final analyses for this paper was de-identified and analyzed in aggregate, with care not to single out any individual nor violate users' privacy. 
Given that Post Guidance rules are not transparent to the user (unless triggered), we chose not to name communities or tie communities to specific rules unless we obtained specific consent from the moderators. We argue that the benefits of this study greatly outweigh its potential harms, given that designing tools to understand online communities better can greatly help improve the Web as a whole.

\bibliographystyle{ACM-Reference-Format}

\bibliography{ref}

\newpage

\appendix

\section{Pilot Experiment Communities}

\label{appendix:communities-list}
\begin{table}[htbp]
\scriptsize

\caption{Descriptive statistics for the 33 communities in the experiment, as measured at the start of the experiment period. DAU captures average daily visitors to the community. Subscribers are logged-in users who have `joined' the community. Weekly Contributions are the total number of posts and comments submitted to the community over 7 days. Age captures the number of years since the community was first created.} 
\label{tab:my-table}
\begin{tabular}{lcccc}
\toprule
\textbf{Topic} & \textbf{DAU} & \textbf{Subscribers} & \textbf{Weekly Contributions} & \textbf{Age (Years)} \\ \midrule
Q\&As & 1M+ & 10M+ & 100K+ & 15 \\ \midrule
Ethics \& Philosophy & 1M+ & 1M-10M & 100K+ & 10 \\ \midrule
Love \& Dating & 100K-1M & 1M-10M & 100K+ & 14 \\ \midrule
Q\&As & 100K-1M & 1M-10M & 10K-100K & 12 \\ \midrule
Action Games & 10K-100K & 1M-10M & 10K-100K & 6 \\ \midrule
Computers \& Hardware & 100K-1M & 1M-10M & 10K-100K & 13 \\ \midrule
Stories \& Confessions & 100K-1M & 1M-10M & 10K-100K & 14 \\ \midrule
Ethics \& Philosophy & 10K-100K & 1M-10M & 10K-100K & 11 \\ \midrule
Mental Health & 100K-1M & 1M-10M & 10K-100K & 15 \\ \midrule
Q\&As & 100K-1M & 1M-10M & 10K-100K & 13 \\ \midrule
Gaming News \& Discussion & 100K-1M & 1M-10M & 10K-100K & 15 \\ \midrule
Books \& Literature & 100K-1M & 10M+ & 10K-100K & 15 \\ \midrule
Stories \& Confessions & 100K-1M & 10M+ & 10K-100K & 11 \\ \midrule
Toys & 10K-100K & 1M-10M & 10K-100K & 15 \\ \midrule
DIY \& Crafts & 10K-100K & 100K-1M & 10K-100K & 15 \\ \midrule
Stories \& Confessions & 10K-100K & 1M-10M & 10K-100K & 9 \\ \midrule
Action Games & 100K-1M & 1M-10M & 10K-100K & 3 \\ \midrule
Makeup & 10K-100K & 1M-10M & 10K-100K & 13 \\ \midrule
Consumer Electronics & 100K-1M & 1M-10M & 10K-100K & 15 \\ \midrule
Comics & 10K-100K & 1M-10M & 10K-100K & 15 \\ \midrule
Space \& Astronomy & 100K-1M & 10M+ & 10K-100K & 15 \\ \midrule
Gaming Consoles \& Gear & 100K-1M & 1M-10M & 0-10K & 7 \\ \midrule
Tattoos \& Piercings & 10K-100K & 100K-1M & 0-10K & 15 \\ \midrule
Consumer Electronics & 100K-1M & 1M-10M & 0-10K & 15 \\ \midrule
Gaming News \& Discussion & 10K-100K & 1M-10M & 0-10K & 11 \\ \midrule
Cars \& Trucks & 10K-100K & 1M-10M & 0-10K & 13 \\ \midrule
Travel \& Holiday & 10K-100K & 1M-10M & 0-10K & 14 \\ \midrule
Science News \& Discussion & 100K-1M & 10M+ & 0-10K & 15 \\ \midrule
Writing & 10K-100K & 10M+ & 0-10K & 13 \\ \midrule
Software \& Apps & 10K-100K & 1M-10M & 0-10K & 15 \\ \midrule
Career & 0-10K & 100K-1M & 0-10K & 15 \\ \midrule
Food \& Recipes & 10K-100K & 1M-10M & 0-10K & 13 \\ \midrule
Filmmaking & 0-10K & 100K-1M & 0-10K & 13 \\
\bottomrule
\end{tabular}

\end{table}


\section{Post Guidance Rules}
\label{post_guidance:rules_examples}

Here, we provide several categories of Post Guidance rules, along with specific (anonymized) examples.

\subsection{Post Length}

\vspace{0mm}

\subsubsection{Character minimums}
\begin{itemize}
    \item Rule name: Character Limit -- 25 Character Minimum
    \item Phrase type: Regex
    \item Regex: \verb!^(.|\s){1,25}$!
    \item Included or Missing: Included
    \item Part of post to check: Post Title and/or body
    \item Message copy: \textit{Your post doesn't meet our minimum character requirement. Please compose a more descriptive post in order to continue.}
    \item Action: Prevent posting.

\end{itemize}

\vspace{2mm}

\subsubsection{Character maximums}

\begin{itemize}
    \item Rule name: Character Limit -- 1000 Character Maximum
    \item Phrase type: Regex
    \item Regex: \verb!^(.|\s){1000}.+!
    \item Included or Missing: Included
    \item Part of post to check: Post body
    \item Message copy: \textit{Your post body exceeds our character limit. Please shorten the length of your post in order to continue.}
    \item Action: Prevent posting.
\end{itemize}

\vspace{2mm}

\subsection{Post Content}

\vspace{0mm}

\subsubsection{Required punctuation}

\begin{itemize}
    \item Rule name: Title must end in a question mark
    \item Phrase type: Regex
    \item Regex: \verb!\? *?$!
    \item Included or Missing: Missing
    \item Part of post to check: Post Title
    \item Message copy: \textit{Your post title must be in form of a question. Please ensure your title ends with a question mark to continue.}
    \item Action: Prevent posting.
\end{itemize}

\vspace{2mm}

\subsubsection{Exclude URLs}
\begin{itemize}
    \item Rule name: No URLs in post title.
    \item Phrase type: Regex
    \item Regex:\verb!(https?:\/\/|www\.)\S+?\.!
    \item Included or Missing: Included
    \item Part of post to check: Post Title
    \item Message copy: \textit{You cannot include a URL in the title.}
    \item Action: Prevent posting.
\end{itemize}

\vspace{2mm}

\subsubsection{Prevent keywords}

\begin{itemize}
    \item Rule name: Tech support is prohibited.
    \item Phrase type: Keywords
    \item Keywords: 
    \texttt{ Help; Broken; Fix; Solution; How do I; What do I; Can't connect; Won't connect; Issues; Problem; Faulty; Unable to; Can't sign in; Won't install; Connection issues; Power cycle; Power cycling; Won't sync; Can't sync; Error; What's wrong; Disconnect; Disconnecting; Disconnection; Lag; Lagging; Artifacts; Transfer; Sign in; Account; Disable; Not working; Won't work; Stuck; Installing; Frozen; Freezes; Glitched; Bugged; Bug; Wi-Fi; Wifi; Internet speed; Slow downloads; Slow download; How to; Issue; Doesn't work; Is it possible; Troubleshooting; Troubleshoot; Remote play quality; Remote play image; Image quality issues}
    
    \item Included or Missing: Included
    \item Part of post to check: Post Title
    \item Message copy: \textit{Asking for tech support in posts is prohibited.}
    \item Action: None.
\end{itemize}

\vspace{2mm}

\subsection{Creative uses of Post Guidance}
\vspace{0mm}

\subsubsection{Conveying subreddit rules}

\begin{itemize}
    \item Rule name: Show message to user when body is between 1 \& 100 characters long.
    \item Phrase type: Regex
    \item Regex:\verb!^.{1,100}$!
    \item Included or Missing: Missing
    \item Part of post to check: Post Body Only
    \item Message copy: \textit{Please note: New users \& those who haven't subscribed to the subreddit, will have their posts held for review.}
    \item Action: None.
\end{itemize}

\vspace{2mm}

\subsubsection{Welcoming users}

\begin{itemize}
    \item Rule name: Welcome message.
    \item Phrase type: Regex
    \item Regex: \verb!^(.|\s){0}$!
    \item Included or Missing: Included
    \item Part of post to check: Post Title Only
    \item Message copy: \textit{Welcome to /r/<anonymized>.}
    \item Action: Show message to user.
\end{itemize}

\newpage

\section{Experiment Outcome Distributions}

\begin{figure}[htbp]
    \centering
    \includegraphics[scale=0.5]{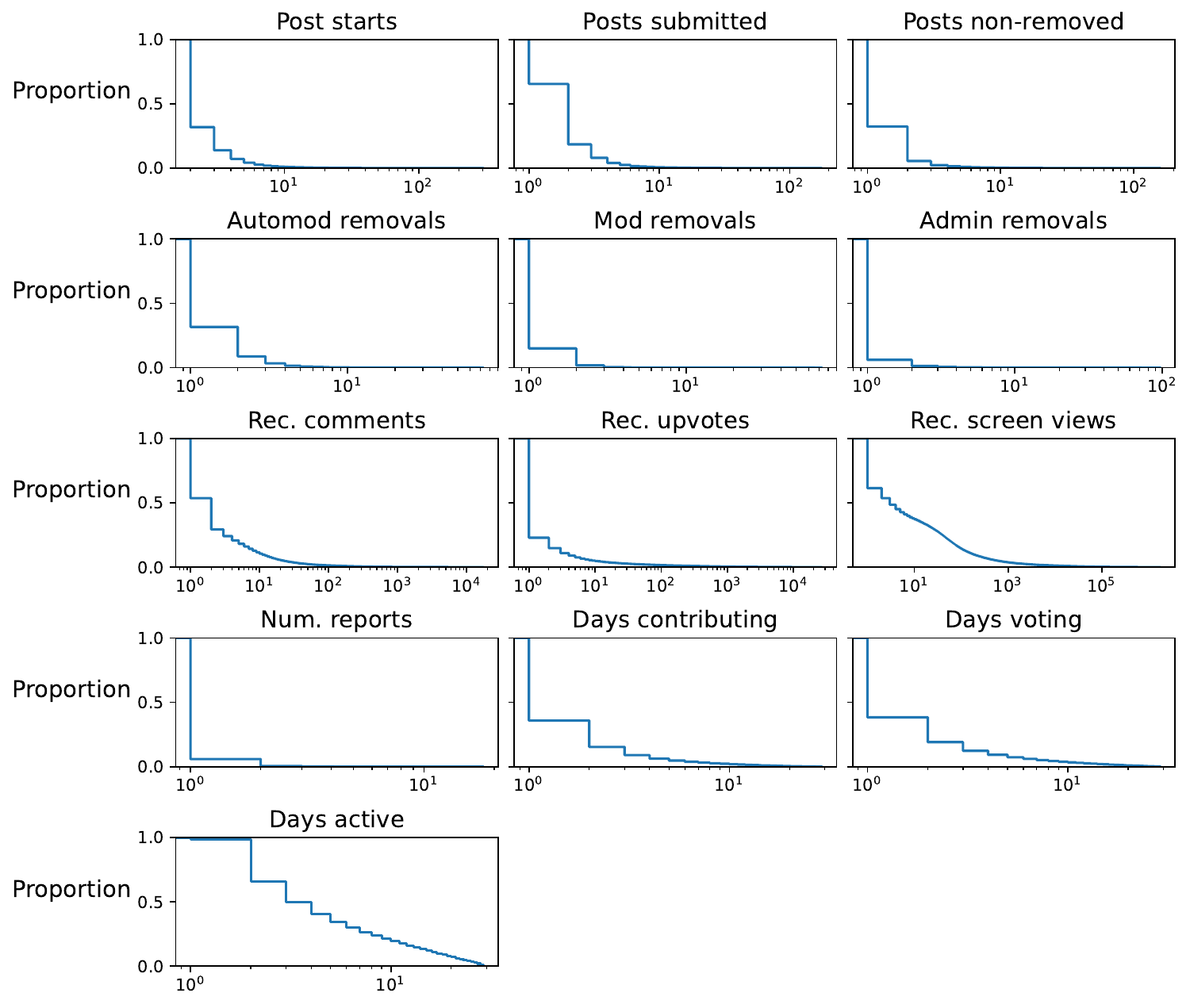}
    \caption{\textbf{Distribution of the outcomes considered in the experiment.} Note that distributions are heavy-tailed, which motivated our choice to model the effect using a Poisson regression.}
    \label{fig:dist}
\end{figure}

\received{January 2024}
\received[revised]{July 2024}
\received[accepted]{October 2024}

\end{document}